# Nonreciprocal cavities and the time-bandwidth limit: comment

Kosmas L. Tsakmakidis[1*], Yun You[2], Tomasz Stefański[3], and Linfang Shen[2*]

[1]Solid State Physics section, Department of Physics, National and Kapodistrian University of Athens, Panepistimioupolis, GR - 157 84, Athens, Greece
[2]Department of Applied Physics, Zhejiang University of Technology, Hangzhou 310023, China
[3]Gdańsk University of Technology, Faculty of Electronics, Telecommunications and Informatics, ul. G. Narutowicza 11/12, 80-233 Gdańsk, Poland

*Equal contribution. Corresponding author's e-mail: ktsakmakidis@phys.uoa.gr

**In their paper in *Optica* 6, 104 (2019), Mann *et al*. claim that linear, time-invariant nonreciprocal structures cannot overcome the time-bandwidth limit, and do not exhibit an advantage over their reciprocal counterparts, specifically with regard to their time-bandwidth performance. In this Comment, we argue that these conclusions are unfounded. On the basis of, both, rigorous full-wave simulations and insightful physical justifications, we explain that the temporal coupled-mode theory, on which Mann *et al*. base their main conclusions, is not suited for the study of nonreciprocal trapped states, and instead direct numerical solutions of Maxwell's equations are required. Based on such an analysis, we show that a nonreciprocal terminated waveguide, resulting in a trapped state, clearly outperforms its reciprocal counterpart, i.e. both the extraordinary time-bandwidth performance and the large field enhancements observed in such modes are a direct consequence of nonreciprocity.**

The paper by Mann *et al*. [1] investigates a time-invariant, unidirectional waveguide interacting with a cavity (Fig. 1a), concluding that the behaviour of the cavity remains unchanged by the presence of the waveguide. This conclusion is then generalised to stating that time-invariant nonreciprocal systems cannot overcome the time-bandwidth (T-B) limit. These assertions appear to conflict with our previous work on a terminated unidirectional waveguide (Fig. 1b), which we deployed to report that large (by a factor of ~1000) T-B violations in linear, time-invariant nonreciprocal systems could be achieved [2].

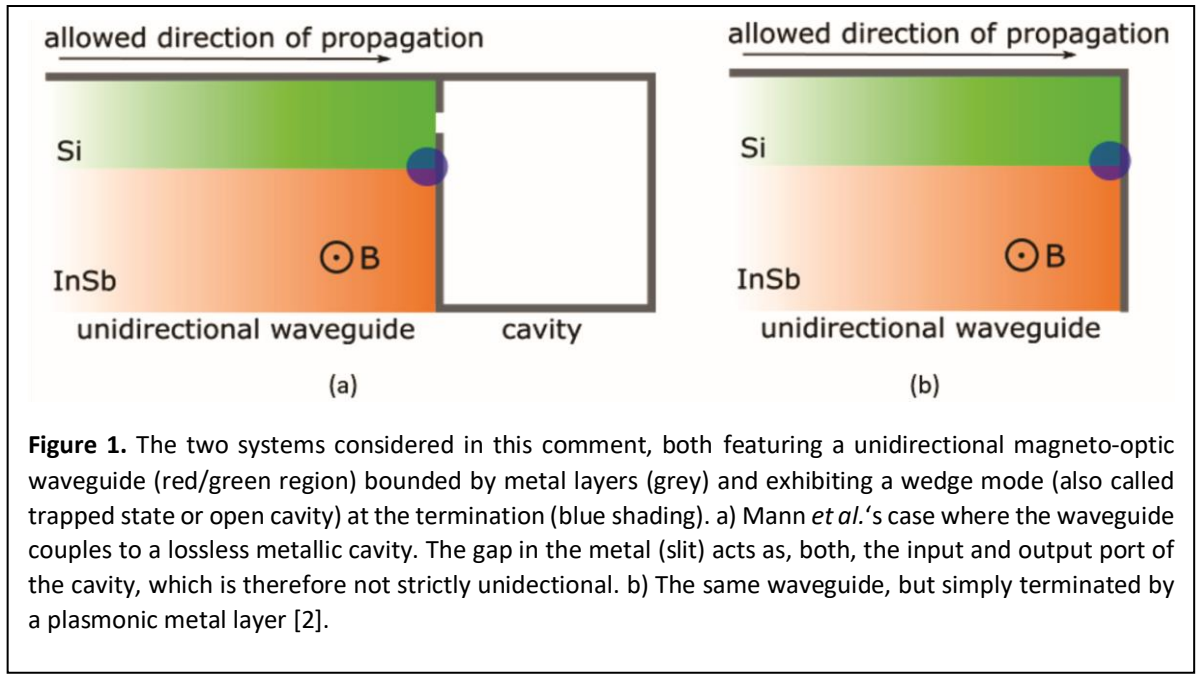

**Figure 1.** The two systems considered in this comment, both featuring a unidirectional magneto-optic waveguide (red/green region) bounded by metal layers (grey) and exhibiting a wedge mode (also called trapped state or open cavity) at the termination (blue shading). a) Mann *et al.*'s case where the waveguide couples to a lossless metallic cavity. The gap in the metal (slit) acts as, both, the input and output port of the cavity, which is therefore not strictly unidectional. b) The same waveguide, but simply terminated by a plasmonic metal layer [2].

However, in this Comment we will show that the discrepancy in the conclusions of these two works stems entirely from the nature of the selected tool of analysis used in Ref. [1]: The main conclusions reached by Mann *et al*. concenring the T-B performance of linear time-invariant systems were on the basis of a temporal coupled-mode theory (TCMT), thereby relying on an analytic *temporal coupled-mode theory approximation* ansatz [1], [3], whereas all the numerical results of Ref. [2] reporting large T-B violations in the same systems were based on full-wave finite-difference time-domain (FDTD) simulations [2]. We will outline that the TCMT used in Ref. [1] is not suited for the study of the structure reported in Ref. [2], which includes a trapped state (blue shading in Fig. 1(b)), and that the extraordinary T-B performance observed in [2] is a direct consequence of the nonreciprocal nature of the device.

To begin, we recall that temporal coupled-mode theory [3] describes the evolution of a field inside a cavity according to the following equation:

$$\frac{da}{dt} = i\omega_0 a - (\gamma_i + \gamma_r)a + \kappa_{in}s_+, \qquad (1)$$

where *a* is the field amplitude inside the cavity, $\omega_0$ the resonance frequency dictated by the cavity (or cavity mode), $\gamma_i$ and $\gamma_r$ the intrinsic and radiative loss rates, respectively, $|s_+|^2$ the power incident onto the cavity from an external system, e.g. a waveguide, and $\kappa_{in}$ the coupling coefficient between that external system and the cavity.

Eq. (1) is satisfied when the field inside the cavity takes the form:

$$a(t) = a_0 e^{i\omega_0 t - t(\gamma_i + \gamma_r)}, \qquad (2)$$

$$\text{where:} \quad a_0 = \frac{\kappa_{in}s_+}{i(\omega_0 - \omega) - \gamma_i - \gamma_r}. \qquad (3)$$

Note, here, that Eq. (3) has the exact same form as Eq. (4) of Ref. [1]. We see that the in-coupling coefficient, $\kappa_{in}$, determines the amplitude of the field inside the cavity; the frequency-dependent behaviour is solely dependent on the denominator. Specifically, the cavity has a half maximum when $|\omega - \omega_0| = \gamma_i + \gamma_r$. Therefore, the bandwidth Δω of an ordinary closed cavity is directly linked to the energy decay rate ($\Delta\omega = \gamma_{tot} = \gamma_i + \gamma_r$) [2], and it is thus inversely proportional to the decay (storage) time, resulting in the well-known time-bandwidth limit [2]. Note that this depends only on the cavity properties, independent of the nature of the feeding waveguide (reciprocal or nonreciprocal).

However, intrinsic to this TCMT description are several key assumptions and approximations, making this approach inapplicable to nonresonant trapped states (e.g., the blue shaded regions in Fig. 1). As shown below, the conclusions drawn by Mann *et al*. based on such a TCMT analysis cannot, therefore, be extended to the trapped state. We note that these states are refered to as 'wedge mode', 'open cavities' or 'trapped states' in varying works, and for the remainder of this Comment we shall use the latter term.

The standard form of temporal coupled-mode theory, analyzed in detail in [1] both for reciprocal and nonreciprocal feeding, assumes that a cavity mode must be a confined, oscillatory mode with a well-defined, single resonance frequency $\omega_0$, that is, it describes resonances peaked at a **single** frequency $\omega_0$ (owing to the '$i\omega_0 a$' term in Eq. (1)). Such an assumption for a resonance, i.e. that it should have a well-defined single peak (at an '$\omega_0$'), though reasonable in ordinary cavities, does not describe key features of the trapped state of Fig. 1b – a point which is now clarified and proved below, with the aid of Fig. 2. Here (in Fig. 2), we apply TCMT to the system [1], [3], [4], with $\omega_0$ being the central frequency of the complete unidirectional propagation (CUP) region [2], and compare these calculations with direct numerical solutions of Maxwell's equations, obtained through FDTD simulations of exactly the same structure and conditions. Specifically, in both cases, we use the same lossy structure (with $\nu = 5\times 10^{-4}\omega_p$ [2], characterizing the losses of InSb) and eliminate cavity back-reflection(s) [3], [4] in both calculation approaches. We see from Fig. 2 that the FDTD calculation predicts a broad and *flat-top* (no single-peak / plateaued) response, while TCMT predicts a narrowband response peaked, as always in that theory, at a single frequency $\omega_0$.

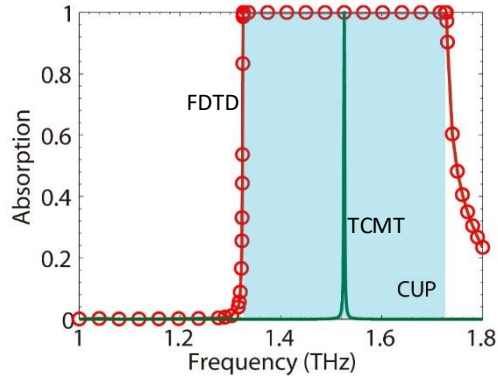

Fig. 2. Absorption resonance of the trapped state in the nonreciprocal structure of Fig. 1b, as calculated through FDTD simulations (red) and a temporal coupled-mode theory analysis (green) – in both cases, for exactly the same lossy structure and reflectionless scenario. From the FDTD simulations it is found that at the center of the CUP region $\gamma_i \sim 3.0614\times 10^9$ rad/s (cf. Eq. (1)). In the temporal coupled-mode theory calculation, the $\gamma_r$ term (cf. Eq. (1)) is also set equal to $\sim 3.0614\times 10^9$ rad/s, to allow for the absence of back-reflection(s) from the terminating cavity [3], [4]. For the FDTD calculation, the shown absorption profile was calculated in the near-field of the excited non-self-sustained [5] bulk plasmon [6] of the terminating (planar) Ag particle, which here itself acts as an open cavity (see also main text), similarly to standard calculations of absorption profiles of plasmonic particles in nanoplasmonics [6].

Which one is the physically correct result? Clearly, the physically correct result is that of the FDTD method because within the CUP region the trapped state cannot radiatively escape its localization region, and therefore it can only 'escape' the system nonradiatively, by eventually been 100% absorbed *within the entire CUP region* – as shown by the red curve in Fig. 2. As such, TCMT is, evidently, not applicable to the trapped state, whose bandwidth is here, as calculated from Fig. 1, in fact ~1000 broader than that predicted from TCMT. This factor (~1000) is in fact the degree to which the T-B limit is overcome in the structure of Fig. 1b, as was reported in Fig. 4 of Ref. [2], precisely because, as was also outlined above, the TCMT always gives rise to T-B-limited resonances [2]. In other words, one cannot deploy (at least, the standard form of) a TCMT approximation, which inherently gives rise to T-B-limited resonaces, to investigate whether a structure might violate (or not) the T-B limit – the result will always be negative, owing to the inherent 'structure' (ansatz) of that theory. For such an analysis, *ab initio* full Maxwell solvers are required, as Fig. 2 above shows, and as Ref. [2] has reported. Note that, interestingly, here, TCMT fails even in the low-loss regime where it is usually successfully applied (e.g., in silicon photonics or in dielectric photonic crystals; cf. lossless structure studied in Ref. [1]), i.e. the failure arises not from the second term on the right-hand side of Eq. (1), but from the first term ('$i\omega_0 a$') on the same side of Eq. (1) – a feature that, to our knowledge, has not been identified in the past, since it does not normally arise in ordinary (non-topological) resonant structures.

Mann *et al*. have also taken the in-/out-coupling rates in a nonreciprocal cavity shown in Fig. 2(a) of Ref. [2] (indicated therein, respectively, with cyan / red colors) to represent the *total* in-/out-coupled energy rate (power), whereas in fact those rates only refer to the *radiative* part of the power, as was explained in Ref. [2] (cf. '$\tau_{out}$' in Fig. 2(a) of Ref. [2] with '$\tau_{out}$' in Eq. (3) of Ref. [2], i.e. the red arrow in both panels of Fig. 2(a) of Ref. [2] is associated with the '$1/\tau_{out}$' *radiative* out-coupling power, not with the total, dissipative '$1/\tau_0$', plus radiative,'$1/\tau_{out}$', rate). In other words, for Lorentz reciprocity to be broken in a cavity resonator, one only needs to (radiatively) in-couple light energy to the cavity, and then the light energy should not *radiatively* escape the cavity – but all light energy will still, nonradiatively, that is via heat, 'escape' the cavity, as it was show in Fig. 3(b) of Ref. [2], and still further herein in Fig. 2. In fact, this is precisely the physical origin of the ~100% absorption in the whole CUP region reported in Fig. 2 herein. We note that this definition of nonreciprocity in a resonator is completely analogous to the well-known definition of nonreciprocity for a waveguide (also reported as Eq. (2) in Ref. [2]) where the wave *transmission* from a point A to a point B should be different from the wave *transmission* from point B to point A – that is, reference is made to the radiative power, to the transmission (we do not 'send' Joule losses from A to B, or from B to A). Thus, to break Lorentz reciprocity in a resonator too, one needs to make unequal only the *radiative* parts of the in-/out-coupled powers – as was reported and explained in Ref. [2]. The total (radiative + dissipative) in- and out-coupled powers are always equal at steady state, as dictated from Poynting's theorem (which is automatically respected in FDTD simulations).

Further, in Ref. [1] Mann *et al*. observe a localized hotspot, which they refer to as a wedge mode, i.e. the trapped state. They conclude that both their trapped state, as well as the one observed in Ref. [2], are not due to nonreciprocity, but simply an example of plasmonic focusing, i.e. a tapered plasmonic waveguide, with nonreciprocity only providing impedance matching. We will now show that this is a misconception, and that nonreciprocity is fundamental to the performance of the device. Specifically, we will show that in the reciprocal version of the device the electromagnetic energy is not confined to a localized region, and while a field enhancement is observed, it does not represent the same focusing nor enhancement factor. Crucially, we will also show that in the reciprocal case the energy of the trapped state decays in a tiny fraction of that of the nonreciprocal structure – that is, the T-B performance of the nonreciprocal structure is drastically superior.

To this end, Fig. 3 reports FDTD calculations, similar to those presented in Ref. [2], displaying, (a), the energy density in the termination as a function of time, and the spatial distribution of the electromagnetic energy at various times for both the reciprocal, (b), and nonreciprocal, (c), case. In all cases, the same device as in [2] is investigated (cf. Fig. 1b), with the difference being that for the reciprocal case no external magnietc field is applied (B = 0 T), while the nonreciprocal case features an applied external magnetic field (B = 0.2 T). Furthermore, the group velocity of the incident light in both cases is almost exactly the same ($v_g$ = 0.0681$c$ for B = 0 T, and $v_g$ = 0.0673$c$ for B = 0.2T), thereby any difference(s) in behaviour cannot be attributed to conventional slow-light effects. From Fig. 3a, we clearly see that the reciprocal device has a much faster decay rate than its nonreciprocal counterpart. In both cases, the pulse propagates (slowly, with the aforementioned group velocities) towards the termination. However, once it reaches the termination the behaviour starts to differ dramatically. For the reciprocal structure, the pulse enters the trapping region and is then back-reflected, resulting in a rapid decay of the energy within the open cavity region. For the nonreciprocal case, however, the backreflection cannot occur (as there is no backwards propagating mode), and light is now trapped in the open cavity region, decaying slowly *only because of dissipative losses* (material absorption). We also note from Fig. 3a that if the magnetic field is reversed (- 0.2 T lines in Fig. 3a), then the pulse can be recovered at times much later than for the reciprocal case. Therefore the nonreciprocal structure clearly outperforms the storage capabilities of conventional plasmonic focusing, with the increased storage time, i.e. delay, being a direct consequence of nonreciprocity.

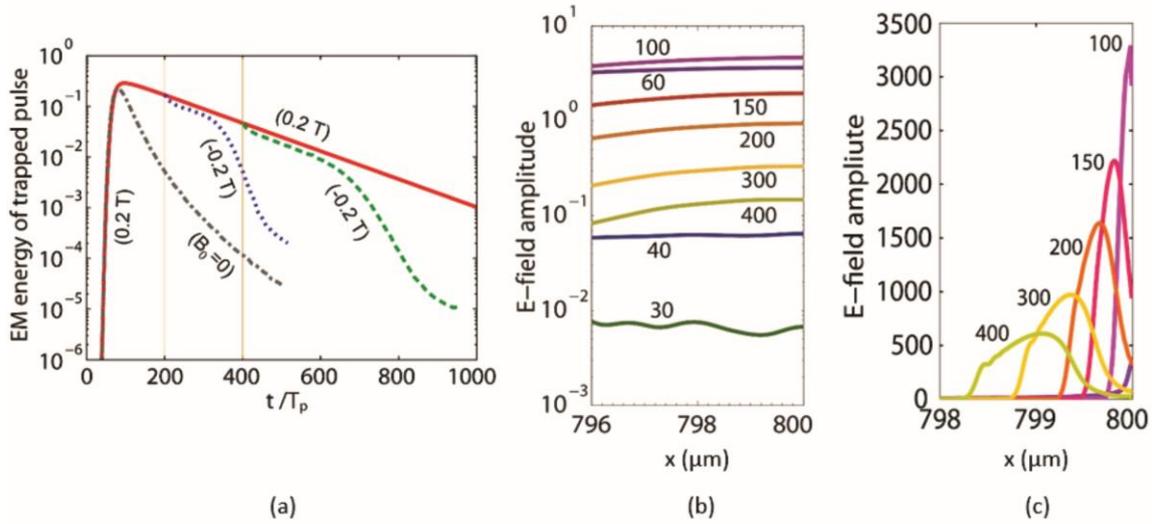

**Fig. 3. Reciprocal vs. nonreciprocal open cavities.** (a) Shown is the time evolution of the electromagnetic energy in an open 'box' adjacently to the terminating end of the structure of Ref. [2], both, for a reciprocal (black dashed-dotted curve; B = 0 T) and nonreciprocal (red solid curve; B = 0.2 T) structure. For the remaining two curves, please refer to Fig. 3(b) of Ref. [2]. (b) Local electric field in the 'box' of the reciprocal structure, recorded at various time instances (in units of $T_p = 1/f_p$, with $f_p$ being the plasma frequency of InSb). (c) Same as in (b), but for the nonreciprocal structure – cf. Fig. 3(a), right panel, of Ref. [2].

To further demonstrate this argument, we show the electromagnetic field within the trapping region (~798 μm < $x$ < 800 μm) for, both, the reciprocal (Fig. 3b) and nonreciprocal (Fig. 3c) cases, at different times, normalized to the amplitude of the incident pulse. We observe that, for both cases, the maximal field enhancement occurs at a time $t$ = 100 $T_p$ ($T_p$=1/$f_p$, where $f_p$ is the plasma frequency of InSb). However, for the reciprocal case we observe only a ~5 times amplitude enhancement (i.e. conventional plasmonic focusing), and we see that over the observed spatial region the field is approximately uniform. In contrast, for the nonreciprocal case we observe that at the same time instant the amplitude enhancement is by a factor of ~3300 – almost 3 orders of magnitude above the conventional result (reciprocal structure). Furthermore, the field is confined in a significantly smaller spatial region. The same pattern is observed at all times, i.e. at any point in time the reciprocal structure has a local field amplitude at the focusing tip several orders of magnitude smaller than that in the nonreciprocal structure, and spreads out uniformly in the spatial region of interest, while the nonreciprocal structure displays extraordinary amplitude enhancement and localization of the field in a smaller spatial region. As such, the argument made in Ref. [1] that the observed effect is conventional plasmonic focusing is clearly unfounded and in contradiction with the observed behavior of the deivce. Both, the field enhancement and T-B performance are dominated by the nonreciprocal nature of the device. These T-B-related differences between nonreciprocal (topological [7, 8]) and reciprocal (ordinary) terminated structures become even more pronounced when realistic surface roughness and material imperfection effects are considered, as it is well-known that reciprocal such structures may even lose their ability to focus and localize light at their tip [9], whereas the nonreciprocal structure of Fig. 1b, being topological [7, 8], is completely immune to such effects [7].

Finally, a few points and clarifications are due with regard to the potential role of nonlocality [10, 11] on the attained, large, T-B violations, as well as on the nature of the 'open cavity' considered in Ref. [2] and in (the blue spot of) Fig. 1b herein: First, the objective of Ref. [2], as well as of the present Comment, was to show that the T-B limit can be exceeded by essentially an arbitrarily high degree in *local* (non-spatially-dispersive), linear, time-invariant structures – that is, the same type of structures considered in Ref. [1], as well as in similar previous works [12-14], which reasoned that no such violations may exist in such structures for fundamental reasons [14]. The results and physical justifications presented here, as well as in Ref. [2], rigorously show that the T-B limit characterizing local, linear, time-invariant structures can be overcome so long as such a violation is *topologically enforced and protected*. Second, even when nonlocal effects are considered, one may always redesign the terminated structure considered here and in Ref. [2], e.g. simply by removing the dielectric (Si) layer, such that it can robustly preserve its unidirectional and topological character even in the presence of nonlocality, and for arbitrarily small levels of dissipation, as has recently been shown in Ref. [11] – thus, nonlocality cannot for fundamental reasons, i.e. for all possible structures, destroy topological protection (topology), since the latter is a deeper and more fundamental property. Third, there is no need for termination and its associated large field-enhancement in a tight region (cf. Fig. 3(c) and brief discussion below), which might give rise to nonlocal effects, as ultrabroadband light trapping [15, 16] and releasing [17] can also exist in topological (unidirectional) 'trapped rainbow' structures [18, 19], which can stretch out and localize (trap) a lightfield in tapered guides in a manner stable even under fabrication disorders [15]. Forth, for device applications of such T-B violations, other important phenomena, such as nonlinear and thermal effects [19], will need to be considered, both of which can however be addressed by e.g. lowering the injected light power or by resorting to cryogenic conditions. It is also to be stressed that the trapped state considered in Ref. [2] and in this Comment, is formed by a non-self-sustained [5] bulk (not surface) plasmon [6] of the terminating Ag layer: The $E_x$-field component, perpendicular to the terminating Ag layer, is dramatically enhanced, inducing free charges on its surface (bulk plasmon), and it is in the near field of that bulk plasmon that the pulse is in-coupled to, without reflection(s) across the entire CUP region. Such plasmonic particles, and their associated bulk plasmons, are typically referred to as 'open cavities' in the field of (nano)plasmonics [6]. Therefore, the lossy topological [7] 'open cavity' (i.e., the Ag particle) in Ref. [2] and herein, is fundamentally different from the lossless perfect-electric-conductor ordinary cavity terminating the unidirectional waveguide of Ref. [1], which therefore, not surprisingly, does not reproduce the behaviour reported in Ref. [2].

In conclusion, the paper by Mann *et al*. [1] makes an interesting contribution in that it convincingly shows that any system whose dynamics are accurately described by (the standard, single-resonance form of) a TCMT approximation, is T-B limited, even when nonreciprocally fed. However, by not recognizing the afore-outlined inherent limitations of such a method of analysis, and the fundamental differences of the structure they considered compared with that in Ref. [2], Ref. [1] reached the generalized conclusion that all (local) linear, time-invariant structures are T-B limited, including the one shown in Fig. 1b herein, studied previously in Ref. [2] – a conclusion which is unwarranted, as explained in some detail above. Moreover, the assertion of Ref. [1] that nonreciprocity does not beget any specific advantage(s) in terms of the T-B performance of a device, is unjustified too, as was clearly shown in Fig. 3 above. Thus, overall, this comment helps to clarify that the time-bandwidth limit *can* be exceeded, in fact to an arbitrarily high degree as Ref. [2] has previously reported, even in (local) linear, time-invariant structures, that topology and nonreciprocity play a crucial role in achieving this feat, and that the standard form of, otherwise powerful, quasi-analytic techniques, such as TCMT that was deployed in Ref. [1], fail to accurately describe the dynamics and physics of (open) nonreciprocal cavities, even in the low-loss regime where they are normally successfully applied.

**Funding.** K.L.T. was supported by the General Secretariat for Research and Technology (GSRT) and the Hellenic Foundation for Research and Innovation (HFRI) under Grant No. 1819.

**Acknowledgment.** The authors would like to thank Sebastian A. Schulz, Jeremy Upham, and Robert W. Boyd for many stimulating discussions and useful comments made on this work. The first of these colleagues, in particular, drew our attention to the key 'i$\omega_0 a$' assumption of standard temporal coupled-mode theory, explained in the main text, and has provided Figure 1.

**Disclosures.** The authors declare no competing financial interests or conflicts of interest.

# Further comments on *Optica* 6, 104 (2019) and on the associated Reply [*Optica* 7, 1102 (2020)]

Kosmas L. Tsakmakidis[1], Yun You[2], Tomasz Stefański[3], and Linfang Shen[2]

[1]Solid State Physics section, Department of Physics, National and Kapodistrian University of Athens, Panepistimioupolis, GR - 157 84, Athens, Greece
[2]Department of Applied Physics, Zhejiang University of Technology, Hangzhou 310023, China
[3]Gdańsk University of Technology, Faculty of Electronics, Telecommunications and Informatics, ul. G. Narutowicza 11/12, 80-233 Gdańsk, Poland

**The pages below provide further results and explanations on the key points made in the main Comment, as well as further elucidating comments on the Reply to the Comment. The pages below have not been peer-reviewed during the peer review of the Comment, and should not be understood as a 'Supplementary' document of the Comment. They could not be part of the Comment in any case, owing to text-length restricitons applied to Comments. Their aim is to further delineate the subtle issues involved in the considered problem, and to clarify unwarranted conclusions articulated in the commented-upon Optica paper (Mann *et al*., Ref. [1] of the Comment) and in the Reply.**

First, in points **i-vi** below we provide a brief summary of all the main comments made in this document. We note that the Reply essentially reiterates almost all of the assertions of the commented-upon Optica paper, without nonetheless disproving any of the actual numerical results (based on full-wave FDTD simulations) of the Comment, but still disagreeing with it. Both, the commented-upon Optica paper and the Reply base their claims and conclusions on the following unwarranted premises:

   **i.**   That the same (prolonged) localization, thereby potential T-B violation, may also occur in tapered plasmonic waveguides. This assertion / statement is not true, and the reason for that is well established in the field of slow light: It completely neglects the role of (tiny, occasionally nm-scale) surface roughnesses, which prevent the attainment of stopped / localized light in tapered waveguides. Until now, in almost two decades, there has not been a single experiment, either in the fields of plasmonic / metamaterial guides, or also in photonic crystal or coupled-resonator optical waveguides (CROWs), demonstrating 'slow light' by a factor of more than a few hundreds, at best, and, most often (if group-velocity dispersion is to be avoided), by a few tens only; nowhere even near light 'stopping' / localization – mainly because of 'innocuous' (tiny) surface roughnesses. This is a well-established fact in the field of 'structural' slow light – among others, outlined in the following Review article: K. L. Tsakmakidis, *et al*., *Science* **358** (6361) –, but is unfortunately completely overlooked in the arguments to that end put forward by Mann *et al.* .

   More broadly, this is one of the main reasons why nonreciprocity, and topology, are needed in photonics: Surface roughnesses, and other material or structural imperfections, lead to back-reflections, and so forth, necessitating e.g. the use of isolators in integrated optoelectronics – even in the 'normal' (not slow) light regime where those devices operate. Thus, if nonreciprocity and topology are indeed needed in that 'normal' light regime in order to avoid such deleterious back-reflection effects, they are even more needed in the slow / stopped light regime, where because of its low group velocity / speed, light now has more time to interact with such roughnesses / imperfections, hence the same problem is only massively exacerbated. Because this crucial factor (the role of surface roughness) is neglected in the argument(s) made by Mann *et al*., their assertions and statements related to Fig. 1 of the Reply are therefore inadvertently misleading. This point is also explained in more detail on pp. 13-14 below, in this document.

ii. That the standard, single resonance, equation describing conventional resonators / cavities leads to T-B limited responses, even when that resonance / resonator is nonreciprocally fed – and that this later point was mentioned/introduced in *Science* **356**, 1260 (2017). This statement is not true: In *Science* **356**, 1260 (2017), we did remind readers of the aforementioned standard equation:

$$\frac{d\alpha}{dt} = i\omega_0 \alpha - \left(\frac{1}{\tau_0} + \frac{1}{\tau_{out}}\right)\alpha + \rho_{in} s_+$$

but, as can clearly be seen in (on p. 2 of) the paper, that was done in order to, first, remind the properties of the equation describing <u>standard/conventional</u> resonators/cavities. The Science paper, then, only writes that:

*However, if Lorentz reciprocity is by some means broken in this passive, linear, and time-invariant resonant system, |ρ$_{in}$| and τ$_{out}$ can become completely decoupled, in which case the product Δωτ (or, equivalently, |ρ$_{in}$|τ$_{out}$) can be engineered at will and take on arbitrarily large values—i.e., in such a case we can exceed the conventional time-bandwidth limit by an arbitrarily large factor.*

In other words, there was no equation provided (no analytic theory was developed) for the case of a *non-reciprocal / unconventional / topological* resonators, simply because if |ρ$_{in}$| and τ$_{out}$ become completely decoupled (as stated in the above-quoted paragraph) it may well happen that the first term too in the RHS of the above equation (the 'i$\omega_0 \alpha$' term) may need to change, or perhaps other modifications of the standard equation might be needed to properly 'capture' the dynamics of the nonreciprocal hotspot (nonreciprocal 'open cavity') – and, in fact, this is precisely what the Comment points out.

Mann *et al*. have misunderstood this subtle point, as a result of which they only investigate various cases / possibilities for the in- / out-coupling coefficients in a 'nonreciprocal resonance / resonator' – but with all cases revolving around the exact same above-quoted equation, with the 'i$\omega_0 \alpha$' term always taken to be as in the standard case. As a result, their theory of 'nonreciprocal resonators / cavities' fails to reproduce the full-wave FDTD results / conclusions of *Science* **356**, 1260 (2017). We stress here that the totality of the conclusions on large T-B violations in *Science* **356**, 1260 (2017) were based exclusively on such full-wave FDTD simulations – i.e., contrary to what Mann *et al*. appear to suggest, inadvertently misrepresenting the Science paper, those conclusions / finding (of the Science paper) were not based on temporal coupled-mode theory calculations at all.

iii. Furthermore, and related to the above, to confront to the standard equation, Mann *et al*. have altered the structure of *Science* **356**, 1260 (2017), using at the end of a nonreciprocal waveguide a standard metallic cavity, which fully confronts to the standard equation and therefore its T-B performance is ordinary – but with this only being a conclusion for that, very standard, structure; i.e., it is not a conclusion that can be generalized, and certainly not a conclusion applying to the localized excitation (hotspot) at the end of the terminated unidirectional waveguide of *Science* **356**, 1260 (2017). That localized excitation (hotspot) is an <u>open</u> resonance, not a standard Lorentzian resonance; therefore, the analysis by Mann *et al*., though mathematically detailed, by focusing on and analyzing the properties of standard Lorentzian resonances when nonreciprocally fed, completely misses the main point – i.e., that it is precisely by <u>not</u> confronting to the standard equation that the localized excitation (nonreciprocal hotspot) can exceed the T-B limit, as was demonstrated rigorously (on the basis of full-wave simulations) in *Science* **356**, 1260 (2017) and in the Comment.

iv. That the unequal in-/out-coupling rates in Fig. 2A of *Science* **356**, 1260 (2017) allegedly imply that the open cavity (hotspot) in the structure of *Science* **356**, 1260 (2017) cannot reach equilibrium / it may be '…*inconsistent with basic thermodynamics…*'. This assertion / statement is completely untrue, arising only from the fact that Mann *et al.* have mistakenly taken the term '$\rho_{out}$' (radiatively out-coupled energy) in Fig. 2A of *Science* **356**, 1260 (2017) to be the total (radiatively + dissipatively) out-coupled energy from the open nonreciprocal cavity / hotspot – this point too is explained in some detail and clarified on pp. 9-11 herein.

v. Related to the above, that the inequality of the in-/out-coupling rates in Fig. 2A of *Science* **356**, 1260 (2017), with '$\rho_{out}$' corresponding to the non-radiatively out-coupled energy, "…*trivially corresponds to a critically coupled cavity. This is a well-known (and completely reciprocal) phenomenon that results in perfect absorption in resonators* …". Here, Mann *et al.* entirely overlook the fact that (perfect) critical coupling occurs <u>only at a single frequency</u> $\omega_0$ – again, this is associated with the '$i\omega_0$' term in the standard, single-resonance ansatz / equation describing Lorentzian resonances. In contrast, the 'critical coupling' (absence of reflection) in the open (nonreciprocal) resonator / hotspot in the structure of *Science* **356**, 1260 (2017) occurs <u>over a very broad</u> and continuous range of frequencies – see also Fig. 2 of the Comment for the difference between the standard critical coupling, and the broadband 'critical coupling' to the nonreciprocal hotspot of the Science paper. Thus, while the standard critical coupling phenomenon is narrowband (hence T-B limited), the reflectionless in-coupling of energy to the localized hotspot in *Science* **356**, 1260 (2017) is very broadband – giving rise to overcoming the T-B limit of not only ordinary resonances but also of any other T-B limit ever written down (in other forms) by orders of magnitude.

vi. Overall, the key point overlooked in *Optica* **6**, 104 (2019) and in the Reply is that the broadband hotspot at the end of the terminated unidirectional waveguide of *Science* **356**, 1260 (2017) remains 'there', localized, that is, delayed, for times $\Delta t$ solely determined by material losses (which can be very small, e.g. assuming cryogenic conditions) and completely decoupled from (as well as much larger than the inverse of) the bandwidth $\Delta \omega$ of the localized excitation – thereby overcoming the standard-resonance T-B limit, or any other T-B limit ever written down, by orders of magnitude. This performance cannot be obtained in (non-topological / standard) tapered plasmonic waveguides, as Mann *et al.* argue, because tiny surface roughnesses completely destroy the localization in the stopped light regime – see point-i above. We are not aware of any other localized wave excitation (hotspot), including those in fields as varied as plasmonic nanoparticles, metamaterial resonators, dielectric nanoparticles, Anderson localization, optical microcavities, and so forth, where a localized excitation decays with a decay rate (determining the time $\Delta t$ over which the excitation is 'stored' / localized) unrelated to its bandwidth $\Delta \omega$ – and that is, in our opinion, a rather extraordinary result, overlooked in the analysis and arguments put forward by Mann *et al.* . Furthermore, this behaviour cannot be theoretically / analytically studied on the basis of the standard, single-resonance, form of temporal coupled-mode theory pursued by Mann *et al*. – as is explained and shown in the Comment using full-wave FDTD calculations.

The above points are now explained in more detail in the following:

To begin with, we note that in the introduction of Ref. [2], Eqs. (3) and (4) were provided in order to explain how the T-B limit arises in ordinary / reciprocal structures [Eq. (3) of Re.[ 2] is the standard, single-resonance, form of the temporal coupled-mode theory approximation]. It was then stated that "… *However, if Lorentz reciprocity is by some means broken in this passive, linear, and time-invariant resonant system, |ρin| and τout can become completely decoupled, in which case the product ΔωΔτ (or, equivalently, |ρin|τout) can be engineered at will and take on arbitrarily large values—i.e., in such a case we can exceed the conventional time-bandwidth limit by an arbitrarily large factor. … .*" In other words, for the T-B limit to be overcome, it is essential that the temporal coupled-mode theory-based analysis of a system is **violated** – not confronted to –, i.e. that the dynamics of the system are **not** accurately 'captured' by the temporal coupled-mode theory approximation, simply because the latter always describes T-B-limited scenarios – as explained, both, in Ref. [2] and Ref. [1]. Thus, we emphasize that the central claim of Ref. [2] – that is, the existence of large T-B violations in nonreciprocal open cavities – was not articulated around Eqs. (3) and (4) of Ref. [2] (i.e., it was not based on temporal coupled-mode theory calculations) but rather *on direct full-wave FDTD solutions of Maxwell's equations* which, essentially, do not involve any approximation(s). The totality of the results shown in Ref. [2], summarized in its Figs. 3 & 4, on which exclusively was based the finding on the existence of large T-B violations in nonreciprocal open cavities, were obtained using such FDTD full-wave simulations, and there is indeed not a single obtained numerical result using the temporal coupled-mode theory approximation in Ref. [2] on which we might have based our aforementioned finding.

In contrast, Mann *et al*. base their central assertions on whether the T-B limit can be broken (or not) in linear time-invariant systems *on a temporal coupled-mode theory analysis* [Eq. (1) of Ref. [1], and its associated analysis detailed throughout]. **On that basis**, following a detailed analysis, Mann *et al*. arrive at two fundamentally different conclusions than Ref. [2]: First, they reason that the T-B limit cannot be overcome in linear time-invariant systems; and second, that nonreciprocal structures / cavities do not exhibit any fundamental advantage over their reciprocal counterparts in terms of (directly and/or indirectly) the T-B limit. In the results and explanations presented in the main Comment and herein, we show that both of these assertions are fundamentally incorrect, and only arise because of the inappropriate nature of the temporal coupled-mode theory approximation to capture / describe the physics of the involved topological resonance associated with the nonreciprocal cavity.

Figure S1 below clarifies the situation: It reports the absorption resonance of the unidirectional cavity of Ref. [2], as calculated from full-wave FDTD simulations (red line with symbols) and from the temporal coupled-mode theory (TCMT) approximation {from Eq. (4) of Ref. [1], as e.g. detailed in Ref. [4]; green solid line}. All calculations were done within the complete unidirectional propagation (CUP) region of the pertaining structure, within which region the nonreciprocal cavity is formed, and for the case where the anisotropic-permittivity loss factor of InSb is $\nu = 5\times10^{-4}\omega_p$. From the FDTD simulations of this structure we find that the absorption rate at the center of the CUP region is ~3.0614×10$^9$ rad/s, which we then use for the $\gamma_i$ term in Eq. (4) of the commented-upon *Optica* paper. Furthermore, to allow for a comparison between the two methods (FDTD vs. TCMT) for exactly the same scenario / case, that is, to allow for absence of reflection from the unidirectional cavity, in the TCMT calculation the out-coupling rate ($\gamma_r$ in the *Optica* paper) is also set equal to ~3.0614×10$^9$ rad/s.

Evidently, there is a dramatic difference between the physically correct FDTD result, based on full-wave direct solutions of Maxwell's equations in the time domain, and the prediction of the TCMT analysis – both, for exactly the same (lossy) structure / scenario (of Ref. [2], with $v = 5\times10^{-4}\omega_p$). In particular, the TCMT analysis completely fails to predict the correct behaviour (shape of the resonance) associated with the trapped pulse ('hotpot') at the end of the terminated unidirectional guide. For the low-loss structure considered here (where $v = 5\times10^{-4}\omega_p$), the TCMT-predicted resonance is extremely narrowband – hence, that theory 'predicts' that the T-B limit is not violated / overcome, as, both, Ref. [2] and Ref. [1] describe for the TCMT 'picture' of analysis. However, as we explain below, this is only an artifact of the approximative nature of that theory (see next paragraph for more details). An analysis that is correct / accurate at the level of Maxwell's equations (FDTD curve) clearly shows that the resonance is, actually, very broadband, covering the entire CUP band. **The degree to which this, physically-correct, FDTD (red) curve is broader than the T-B-limited TCMT (green) curve, is precisely the degree to which the T-B limit is exceeded in this structure** – exactly as was reported in Fig. 4(b) of Ref. [2], namely by a factor of ~1000 (!). We also note that **this overcoming of the T-B limit occurs within a physically very tight ('box') region, that is, the T-B limit is exceeded within a (very) small 'footprint', rather than trivially, e.g. merely by increasing the length of the guide.** Thus, it is clear that the central conclusion of Ref. [1], namely that the T-B limit cannot be overcome in this structure, based on an analysis of the TCMT equation and its associated properties, is unwarranted – i.e., even though the therein-presented TCMT-based analysis is mathematically correct, the aforementioned central assertion of the commented-upon *Optica* paper is incorrect, because in this situation the TCMT approximation fails to accurately describe the physics and dynamics of the unidirectional cavity (where the pulse is trapped / localized, at the end), hence any assertion or conclusion based on it, particularly with regard to whether the T-B limit might be exceeded or not, is ill-founded.

Why is there such a discrepancy between the FDTD- and TCMT-predicted results (for exactly the same structure and conditions) shown in Fig. S1(a)? There is a precise physical reason for that. In the unidirectional cavity of (Fig. 2(b) of) Ref. [2], all the injected / in-coupled (into the cavity) light-pulse energy is completely (that is, 100%) absorbed, as long as the light-pulse's bandwidth is within the CUP region – simply because the therein-trapped light-pulse's energy has nowhere else to 'go' (e.g., it cannot be back-reflected, nor scattered in the bulk InSb modes, etc); hence, all of it, is eventually absorbed, i.e. it 'escapes' the system nonradiatively as heat. Importantly, <u>**this happens for any degree of losses**</u> (the '$v$' parameter mentioned above) of the InSb, that is, even when, say, $v = 5\times10^{-4}\omega_p$, or when $v = 5\times10^{-3}\omega_p$, or when $v = 5\times10^{-2}\omega_p$, etc; in all cases, the absorption within the CUP region is always 100% – the only thing that changes each time is the **rate** with which the trapped-pulse energy is absorbed, that is, for lower losses (lower values of $v$) it takes a longer time for the trapped pulse to be completely absorbed. This is precisely what was explained, e.g., in the introduction of Ref. [2] quoted above on p. 1 herein. In other words, in this terminated unidirectional structure, the absorption of the trapped pulse is always 100% within the CUP region, **irrespective of the material losses of InSb**. Different material losses of InSb, only change the **rate** with which the trapped pulse is absorbed; smaller InSb material losses lead to smaller decay / absorption rate of the trapped pulse, while higher losses lead to higher rates – but in all cases, all (100%) of the trapped-pulse's energy is eventually absorbed within the same, constant, bandwidth of the CUP region; it is just that for lower InSb losses it takes longer time for the trapped pulse to be completely absorbed. Thus, **in this system the decay rate (which determines the storage time of the system) is completely decoupled from the bandwidth of the system (which is always constant; the CUP bandwidth)** – hence the T-B limit can indeed be exceeded, even, by orders of magnitude, exactly as was reasoned and shown (on the basis of full-wave FDTD simulations) in Ref. [2]. We also note that the physically-correct, FDTD-predicted, result of Fig. S1(a) (red curve with symbols), predicts a resonance which is **very 'flat'** within the CUP region, that is, **it does not exhibit an isolated peak at an individual / isolated frequency point** – it remains unity (100%) for all frequencies within the CUP region, for the reason(s) just explained above.

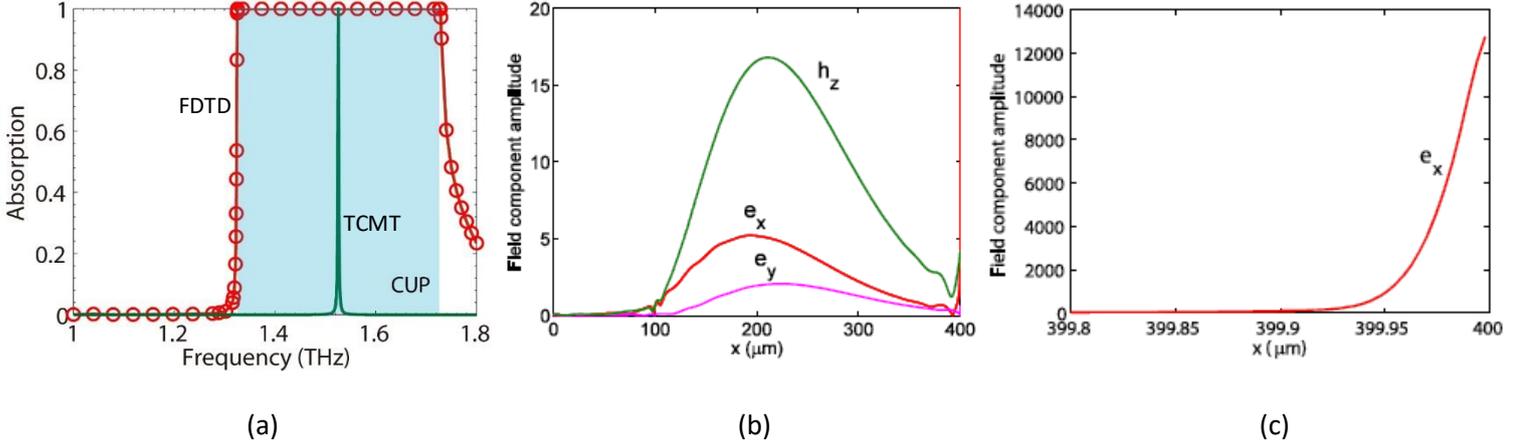

**Fig. S1. Characteristics of the nonreciprocal guide and cavity.** (a) Absorption resonance of the nonreciprocal cavity, within the CUP frequency region, as calculated with the full-wave FDTD method (red solid line with symbols) and – for exactly the same structure and conditions – with the temporal coupled-mode theory (TCMT) approximation. For the FDTD calculation, the shown absorption profile was calculated in the near-field of the excited non-self-sustained [5] **bulk** plasmon of the terminating (planar) Ag particle (which here itself acts as an open cavity – see also point-1 on p. 5 below), similarly to standard calculations of absorption profiles of plasmonic particles in nanoplasmonics [6]. (b) Snapshot of the (amplitudes of the) three field components along the InSb surface before the guided surface magnetoplasmon reaches the terminating Ag layer. The amplitudes of the *E*-field components are comparable to (in fact, slightly smaller than the) amplitude of the *H*-field component. (c) Snapshot of the (amplitude of the) $E_x$-field component (i.e., the *E*-field component perpendicular to the terminating Ag layer) when the guided pulse of (b) has now been in-coupled, without reflection(s) over the entire CUP band, to the extreme near-field of the bulk plasmon of the terminating plasmonic Ag 'particle' / layer. Note the scales of, both, the horizontal and vertical axes. The other two field components are not shown, for clarity, as they are orders of magnitude smaller, but their maximum values are mentioned in the text.

This contrasts sharply with the TCMT prediction(s). One may readily infer that, even from a simple look at the (standard, single-resonance form of the) TCMT equation (Eq. (3) of Ref. [2], and Eq. (1) of the commented-upon *Optica* paper):

$$\frac{da}{dt} = i\omega_0 a - (\gamma_i + \gamma_r)a + \kappa_{\text{in}} s_+. \qquad (1)$$

The key point is that this equation always describes a single-peak (Lorentzian) resonance, peaked at a single frequency '$\omega_0$', owing to the '$i\omega_0\alpha$' term in it – see also Eqs. (1)-(4) in Ref. [3]. This is clearly shown with the TCMT (green) curve in Fig. S1(a) above. Thus, fundamentally, this equation can never be a good 'fit' to the flat-peak actual resonance (FDTD / red curve) describing the system. Even more so, for very small losses (the $\gamma_i$, $\gamma_r$ terms in Eq. (1)), this equation, describing a Lorentzian response, gives rise to a correspondingly narrowband Lorentzian resonance, exactly as was described in the introduction of Ref. [2]. Thus, this equation, as is also clearly shown in Fig. S1(a) above, fundamentally underestimates the actual bandwidth of the nonreciprocal cavity. Hence, any assertion based on it, particularly with regard to whether a nonreciprocal cavity overcomes the T-B limit or not, is fundamentally ill-founded. Unfortunately, Mann *et al.* analyze (in Ref. [1]) in some detail the properties of just such a TCMT equation, and – on that, as it turns out, inappropriate basis – concluded that the T-B limit cannot be overcome in such a system – a conclusion which is, thus, groundless, i.e. an artifact of the inappropriate nature of the TCMT for the present problem. This is also the reason that the totality of the presented results / calculations shown in Ref. [2] were based on full-wave FDTD simulations, solving directly Maxwell's equations in the time-domain, rather than on a TCMT analysis / approximation – and it was entirely on that basis (i.e., on the rigorously obtained FDTD calculations) that it was therein reported that large T-B violations exist, in particular in Fig. 4 of Ref. [2].

Having reached the aforementioned central, but unfortunately incorrect, conclusion on whether the T-B limit can be exceeded in local, linear, time-invariant structures, owing to the inappropriate basis (the TCMT approximation) on which they investigated the problem at hand, Mann *et al.* then, unfortunately, went on to make a series of additional unwarranted statements, as well as reach further incorrect conclusions, perhaps attempting to justify their above assertion. We highlight, explain, and clarify these in the following:

1) The first, concerns the existence of a 'nonreciprocal cavity'. Where exactly is such a cavity, e.g. in Fig. 2(b) of Ref. [2]? Figures S1(b), (c) above, clarify the situation. In particular, one may see from Fig. S1(b) that initially, while the surface magnetoplasmon (SMP) is guided along the one-way structure (without still having reached its terminating end), the guided wave is in the full 'electromagnetic / electrodynamic regime' where all three field components are comparable in amplitude. The situation changes drastically when the wave reaches the terminating end: In that region, the $E_x$-field component (the *E*-field component perpendicular to the terminating Ag layer) is dramatically enhanced, by a factor of ~2800, as shown in Fig. S1(c), while the other two field components attain maximum values of $E_y$ ~140 and $H_z$ ~50 (not shown in Fig. S1, owing to the differences in the scale). In other words, in that region one has $|\vec{E}| \gg |\vec{H}|$, i.e. we enter the electro-quasi-static regime, and there is a large field discontinuity in the *E*-field component perpendicular to the terminating Ag layer, that is, there is a large concentration of electric charges on the surface of the Ag layer. This means that the non-self-sustained [5] bulk (not surface) plasmon of the terminating Ag plasmonic layer is excited. Indeed, one may note from (the horizontal axis of) Fig. S1(c) how extremely tightly the $E_x$-field component exists to the surface of the terminating Ag plasmonic layer. The situation is completely analogous to the excitation of bulk plasmons in (e.g., spherical) plasmonic particles – widely referred to in the pertinent literature as 'open cavities' [6] –, only that here the Ag 'particle' is planar rather than e.g. spherical or of some other shape. Thus, it is in the extreme near field of this plasmonic (Ag) 'open cavity' that the guided, incident, SMP is 'critically' (that is, without back- or other reflections) in-coupled to, over the whole bandwidth of the CUP region, leading to a dramatic transition from the full electrodynamic regime to the electro-quasi-static regime, with its ensuing $E_x$-field-only drastic enhancement. Note also that, in that region, the trapped wave only decays with time (in a given region of space) owing to dissipative losses, that is, there is not an $e^{ikx}$ term describing its dynamics (i.e., there is no propagation any more along the *x* direction; the structure does not behave as a 'waveguide' any more), i.e. the pulse is spatially localized in the extreme near field of the Ag 'particle' which behaves as a (nonreciprocal open) cavity.

In Ref. [1], this subtle, but key, phenomenon has, evidently, been completely misunderstood, and instead an entirely different structure was introduced / suggested, where a standard metallic cavity was added at the end of the guide (Fig. 2(a) of Ref. [1]) – in an apparent effort to investigate whether such a (standard) cavity might be 'nonreciprocal' as in Ref. [2]. Furthermore, that cavity in Ref. [1] is completely lossless, made of a perfect electric conductor (PEC) – when it is in fact known that, fundamentally, there can never exist a lossless nonreciprocal cavity, simply because in such a system energy balance cannot be established (at steady state, from Poynting's theorem, the in-coupled power must be exactly equal to the out-coupling / escaping power; if the system is lossless, and unidirectional / without back-reflection out-coupling, such a power balance can, hence, never be established). Additionally, the cavity in Ref. [1] is not impedance-matched to the in-coupling waveguide at even a single frequency of the CUP region, and light energy flows, both, into and out of the slit / opening of the lossless PEC cavity, i.e. the cavity is not nonreciprocal. Hence, for that system, unsurprisingly, Ref. [1] did not observe any T-B violation at all, but this conclusion is nonetheless then casually generalized to all 'nonreciprocal cavities' – including the afore-described one of Ref. [2], whose physics are fundamentally different, as shown and explained above. This casual generalization is, thus, unfounded.

2) The second point is whether nonreciprocity plays a role, or not, in overcoming the T-B limit. In the main Comment, but also on point-5 herein, we provide FDTD simulation results, as well as physical insights, clarifications and explanations, showing and proving unambiguously that the nonreciprocal structure has a dramatic advantage – indeed, by orders of magnitude – over its reciprocal counterpart, specifically in terms of all direct and/or indirect figures-of-merit associated with (overcoming) the T-B limit, and especially in realistic experimental situations (presence of surface roughness, defects, inhomogeneities, material imperfections, etc). The fundamental reasons for this are two: First, that the nonreciprocal structure is one-way (that is, it suppresses the back-reflection / 'radiation' loss channel of the trapped pulse), leading to much greater 'storage' lifetimes of the trapped pulse (hence, the time-bandwidth product is greatly enhanced); and second, the nonreciprocal structure is, in fact, topological (rather than, e.g., merely 'magneto-optic' or 'bianisotropic', etc), immune to material or interface imperfections (which often prevent an ordinary / reciprocal structure from focusing a wave at its terminating tip, i.e. from trapping a pulse at its end), and with deep topological reasons (specifically, the breakdown of the 'bulk-edge correspondence') being responsible for its light-trapping performance, unlike ordinary / reciprocal structures. We explain all of these points in some detail on point-5 below.

3) A third misunderstanding in Ref. [1] concerns the in-/out-coupling rates in the nonreciprocal cavity – as was, e.g., schematically illustrated in Fig. 2(a) of Ref. [2], which (for further convenience) is shown below together with the corresponding TCMT equation (Eq. (3)) of Ref. [2]. The authors of the *Optica* paper have mistakenly understood the in-/out-coupling rates in a nonreciprocal cavity shown in that figure (indicated, respectively, with cyan / red colors) to represent the total in-/out-coupled energy rate (power), whereas in fact they (the rates) only refer to the radiative part of the power. This is clearly shown / explained in the figure below (Fig. 2(a) of Ref. [2], together with Eq. (3) of Ref. [2]) where it is highlighted that the red arrow in both panels is associated with the '$1/\tau_{out}$' radiative out-coupling power – not with the total (dissipative, '$1/\tau_0$' + radiative, '$1/\tau_{out}$') out-coupling power. In other words, what this schematic clearly shows is that for Lorentz reciprocity to be broken in a cavity/resonator, one only needs to (radiatively) inject/in-couple light energy to the cavity, and then no light energy should radiatively escape the cavity – but all light energy will nonradiatively, that is, via heat, 'escape' the cavity, exactly as was show in Fig. 3(b) (red curve) of Ref. [2], and still further herein in Fig. S1(a). In fact, this is precisely the physical origin / justification of the 100% absorption in the whole CUP region, reported e.g. in Fig. S1(a) herein. Note that this definition of nonreciprocity in a resonator is completely analogous to the well-known definition of nonreciprocity for a waveguide (also reported as Eq. (2) of Ref. [2]) where the wave *transmission* from a point A to a point B should be different from the wave *transmission* from point B to point A – that is, reference is made to the radiative power / to the transmission (we do not 'send' Joule losses from A to B, or from B to A). Thus, in a resonator too, to break Lorentz reciprocity, the radiative-parts-only of the in-/out-coupled powers should be unequal / asymmetric – as was reported / explained in Ref. [2].

Unfortunately, Ref. [1] has misunderstood this important point, too, taking the out-coupling rates to be associated with the total out-coupled power – as is e.g. illustrated in Fig. 1(a) of Ref. [1]. It then makes a range of unwarranted statements about Ref. [2], e.g. even stating in its introduction that the afore-described inequality of the (radiative) rates "…*raises concerns regarding its thermodynamic validity: for example, a system with unequal input and output rates violates the second law of thermodynamics*. …". Ref. [1] then proceeds by making a detailed analysis to establish what is rather

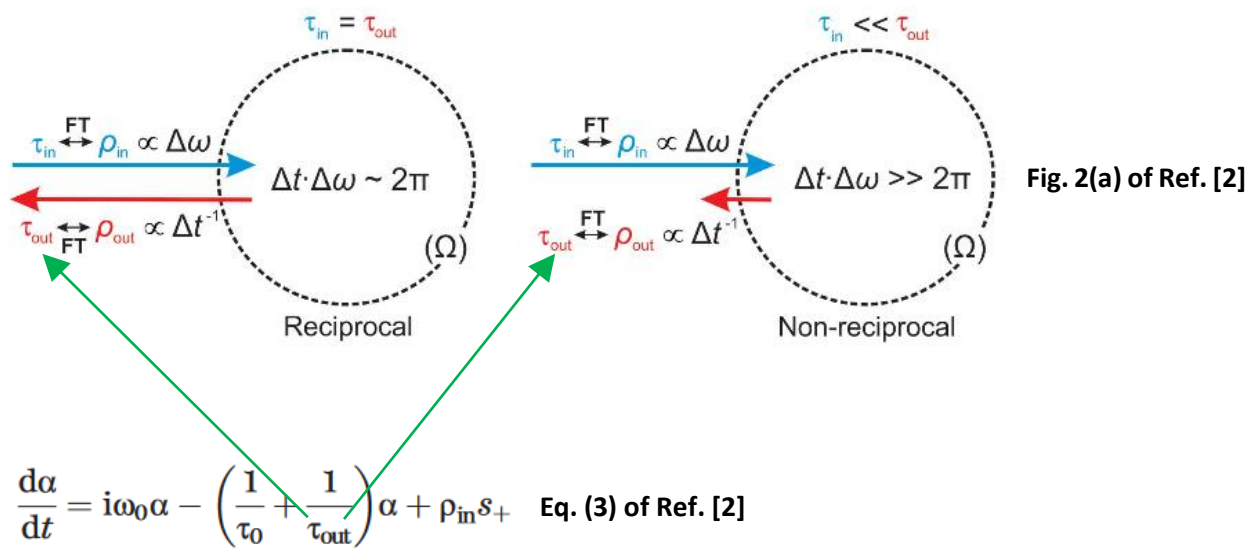

Fig. 2(a) of Ref. [2]

Eq. (3) of Ref. [2]

obvious, that is, power balance at steady state (equal in-/coupling **total** rates; Eqs. (5f) and (S21) of Ref. [1]), which simply follows from Poynting's theorem – and was therefore fully respected in the FDTD simulation results and conclusions reported in Ref. [2]. In doing so, Ref. [1] evidently misunderstood some of the key physics / processes that here enable the T-B violation: That all of the pulse's energy, within a **constant** CUP **bandwidth**, is completely / 100% absorbed (radiative in-coupled power = nonradiative out-coupled power, respecting Poynting's theorem), **irrespective of the material (InSb/Ag) losses**, leading to correspondingly (for different material losses) different pulse decay / absorption rates, hence to different 'storage' times – i.e., that the (CUP) bandwidth and storage times in this nonreciprocal cavity are completely decoupled, leading to arbitrarily large T-B violations, as was explained in detail in Ref. [2] but also in Fig. S1(a) herein.

**4) A forth point concerns another statement (on p. 4) of Ref. [1], namely that:** "… *Aside from the wedge mode in this terminated unidirectional waveguide, similar singularities that support finite absorption in the limit of vanishing material loss can be found in other extreme, yet reciprocal, electromagnetic systems, and are a reminder that ideal lossless scenarios should be generally considered an artifact in electromagnetics and may sometimes lead to singularities and non-unique solutions, especially with negative permittivities*. …" In other words, it is therein implied that 'lossless scenarios' could be the reason for the reported T-B violations in Ref. [2] (which also featured a structure with 'negative permittivities'; the InSb, Ag layers), but that those violations are restored once losses are properly accounted for. As explained and shown above, not a single structure considered in Ref. [2] was lossless, and in fact (dissipative) losses (of the InSb, Ag layers), in all cases, played a key role in enabling the trapped-pulse's energy to nonradiatively 'escape' the system, as heat, thereby establishing power balance. This was clearly shown in, e.g., Figs. 3(b), 4(a) and 4(b) of Ref. [2], showing the nonradiative decay of the trapped pulse, as well as the non-zero values of the loss-parameter '*v*' (quantifying losses in the InSb layer) used in the totality of the therein-reported FDTD simulation results. In contrast, there is only one completely lossless structure used / analyzed in the present problem, and that is the perfect-electric-conductor (PEC) cavity considered in Ref. [1].

For all of the above reasons, but also for the additional ones detailed in our replies to Reviewer #1 below (where we elucidate the key role played by nonreciprocity in obtaining large T-B violations in linear, passive, time-invariant structures, as reported in the *Science* paper), we feel that it would be important and of keen interest to *Optica*'s broad readership that the afore-detailed fundamental objections to the paper by Mann *et al*., which we justify, both, by further explaining and analyzing the physics of the present problem but also on the basis of additionally/newly obtained rigorous numerical calculations, are published as a 'comment' on that article – and we would like to thank the referee for the comments provided, which have prompted us to clarify the important issues that she/he raised.

**5) Role of nonreciprocity / topology:** In the main Comment, and here (below), we demonstrate that a nonreciprocal cavity fundamentally outperforms a reciprocal cavity, specifically in terms of the attained time-bandwidth performance – and/or in relation to any further pertinent key figure of merit.

Figure S2(a) below reports such a result, obtained from full-wave FDTD simulations of exactly the structure investigated in Ref. [2]. The solid red curve in this figure corresponds to the case where a static magnetic field ('bias') of 0.2 T is applied to the structure, making it nonreciprocal (i.e., opening a unidirectional / one-way bandgap), whereas the dashed-dotted black curve corresponds to the same structure but with the applied magnetic field being B = 0, making it an ordinary, reciprocal open cavity (owing to the terminating Ag layer at the end, used in both structures, as in Ref. [2]). Plotted in this figure is the energy of the trapped pulse, in a small 'box' (1 μm × 1 μm) exactly at the terminating area, encompassing Si, InSb and tangential to the terminating Ag layer (the trapped pulse in Fig. 2b of Ref. [2], or the wedge mode in Fig. 2e of Ref. [1]). To facilitate a fair comparison between the two structures – the nonreciprocal open cavity (B = 0.2 T) and the reciprocal open cavity (B = 0 T) – the group velocities of the pulses in both structures are, also, almost exactly the same ($v_g$ = 0.0681$c$ for B = 0 T, and $v_g$ = 0.0673$c$ for B = 0.2 T).

We note the, rather dramatic, difference between these two curves/cases. For the case where B = 0 T (reciprocal structure), the pulse propagates slowly (with a group velocity $v_g$ = 0.0681$c$), reaches/enters the terminating 'box', is back-reflected from the terminating Ag layer, and then slowly (with the same group velocity) exits the box in the opposite direction to the one it entered the box. During this process, as one may discern from the black dashed-dotted curve in Fig. S2(a), the energy in the box initially increases and then quite rapidly decreases / drops off – hence, the storage time of this open reciprocal box is not large, i.e. the structure is (storage-)time-bandwidth limited, similarly to standard/ordinary structures. In contrast, for the nonreciprocal case (red solid curve, B = 0.2 T), the decay rate of the pulse/light energy in the box is much slower; for instance, for the shown-in-Fig.S2(a) energy to decay to $10^{-3}$ it takes ~250 $T_p$ for the reciprocal cavity ($T_p = 1/f_p$ being the inverse plasma frequency of InSb) as compared to ~1000 $T_p$ for the nonreciprocal cavity – with the difference between the storage times of the two cavities increasing even more for larger decays of the pulse energy, owing to the different slope of the two curves. Physically, this simply happens because in the nonreciprocal case, the pulse propagates slowly (with a group velocity $v_g$ = 0.0673$c$), reaches/enters the terminating 'box', and is **not** back-reflected from the terminating Ag layer (owing to the unidirectionality), i.e. it 'stays there', inside the box, for much longer times, limited only by dissipative losses (the reflection/'radiative' losses from the box have been completely eliminated, in sharp contrast to the reciprocal case). These results, therefore, clearly illustrate the fundamental advantage that nonreciprocal cavities have over their reciprocal counterparts, specifically in terms of the attained time-bandwidth performance(s).

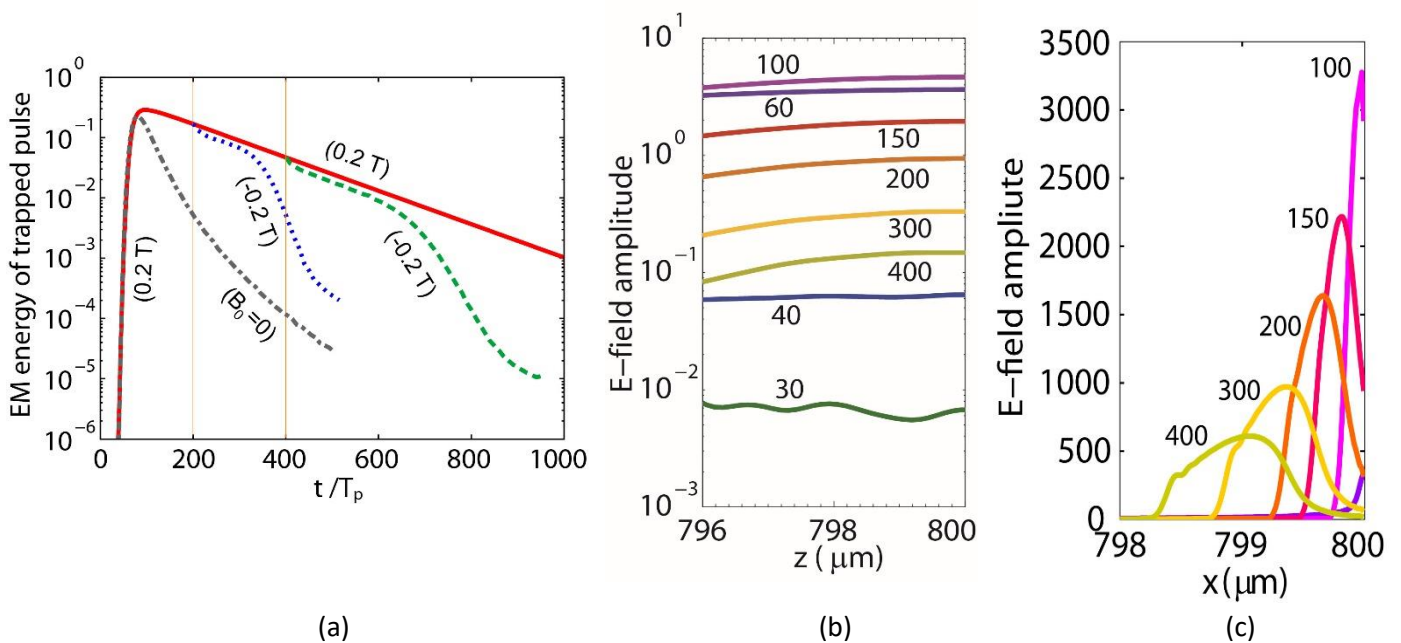

(a)  (b)  (c)

**Fig. S2. Reciprocal (ordinary) vs. nonreciprocal (topological) open cavities.** (a) Shown is the time evolution of the EM energy in a 'box' adjacently to the terminating end of the structure of Ref. [2], both, for a reciprocal (dashed-dotted curve; B = 0 T) and nonreciprocal (red solid curve; B = 0.2 T) structure. For the remaining two curves, please refer to Fig. 3(b) of Ref. [2]. (b) Local electric field in the 'box' of the reciprocal structure, recorded at various time instances (in units of $T_p = 1/f_p$, with $f_p$ being the plasma frequency of InSb). (c) Same as in (b), but for the nonreciprocal structure – cf. Fig. 3(a), right panel, of the *Science* 2017 paper.

Furthermore, and equally importantly, it is to be stressed that the aforementioned fundamental difference between the reciprocal and nonreciprocal structures is not only restricted to the computed values of the time-bandwidth product(s) per se – characterizing the storage capacity of the device(s). Structures that overcome the conventional/standard time-bandwidth (T-B) limit do so, also, in order to enhance the interaction between light and matter, owing to the enhanced interaction/storage time and local-field enhancement (for a given bandwidth) that such an overcoming (of the T-B limit) implies.

To this end, shown in the main Comment and in Fig. S2(c) herein, on the basis of full-wave FDTD simulations directly solving in the time domain Maxwell's equations (i.e., rather than deploying the temporal coupled-mode theory *approximation*), are snapshots at various time instances of the pulse dynamics near the terminating end. We note in all cases, the dramatic difference – by more than two or even three orders of magnitude – between the attained local *E*-field amplitudes in the reciprocal and nonreciprocal cavities/boxes. For instance, at *t* = 100 $T_p$, the local *E*-field amplitude in the reciprocal box is ~5, whereas in the nonreciprocal box it is ~3300 – larger than the reciprocal case by a factor of ~660. At *t* = 400 $T_p$, the local *E*-field amplitude in the reciprocal box is ~$10^{-1}$ whereas in the nonreciprocal box it is ~600 – larger than the reciprocal case by a factor of ~6000.

Thus, owing to its unidirectional character and the associated absence of back-reflection at the terminating Ag layer, a nonreciprocal structure can concentrate the same amount of energy in a much tighter spatial region, giving rise to local *E*-field enhancements that can be *larger by orders of magnitude* compared to those attained by its reciprocal counterpart – a property which could, therefore, be used for such important applications requiring strong light-matter interactions as enhancement of nonlinear effects, sensing, enhanced spontaneous emission rates, and so forth.

Furthermore, in the following, for the sake of outmost clarity, we additionally wish to explain that the differences between the two classes of devices (reciprocal and nonreciprocal) are in fact even deeper, and become particularly important in very practical / realistic situations, again specifically in terms of the attained T-B performances of these two classes of devices.

To begin with, we first note that the one-way nonreciprocal structure used in Ref. [2] is not just 'magneto-optic' (or merely 'bianisotropic') – it is actually a *topological* structure, as has recently been established in a series of pertinent works, e.g. Ref. [7]. Specifically, all the properties that quantify photonic topological insulators (PTI), such as the Berry phase, Berry connection, and Chern numbers, can properly and rigorously be defined/derived in such a structure (as in Ref. [2]) too. There is a key difference though: Because this structure (of Ref. [2]) is made of *continuous* (for photons) media, rather than of periodic (for electrons) media as in topological-insulator semiconductors or in topological photonic crystals (for photons), one needs to introduce a spatial cutoff $k_{max}$ in the pertinent wavevector-domain calculations for the aforementioned quantities (Berry phase, Berry connection and Chern numbers) to be properly/ correctly calculated/defined (Ref. [7]). Practically, this means that between the InSb layer in Ref. [2], and the enclosing upper Ag layer, there needs to be a distance/gap *d* separating them, so that $k_{max} \sim 1/d$. Under this condition, the well-known 'bulk-edge correspondence' can be used to predict the number of unidirectional edge states on the surface of InSb (a medium with no intrinsic periodicity for photons). In Ref. [2], just such a distance *d* was separating InSb with the upper Ag layer, filled with Si.

There are, now, two important points we wish to make/elucidate based on the above concise summary:

No such distance/gap *d* exists between InSb and Ag at the terminating right end of the structure of Ref. [2], that is, InSb directly 'touches' the terminating Ag layer (effectively, *d* = 0 or $k_{max} \rightarrow \infty$). As a result, the 'bulk-edge correspondence' breaks down – as was recently shown, and explained in some detail, in Ref. [7]. This means that there is no surface wave supported at the terminating and lower InSb/Ag interface, through which the localized pulse at the terminated end of our structure might have escaped. In other words, in addition to unidirectionality, there is a rather beautiful topological argument – the breakdown of the bulk-edge correspondence; a phenomenon that does not usually occur in electronic or photonic topological insulators – explaining the absence of 'radiative' (escaping) losses for the localized pulse in our unidirectional cavity. No such effect, of course, exists for the reciprocal/ordinary structure/cavity (B = 0 T), which – as explained above – does exhibit significant radiative losses. As a result, there is a deep topological reason for the enhanced storage time characterizing the nonreciprocal / topological cavity as compared to its reciprocal / ordinary counterpart – with this enhanced storage time being directly linked to the enhanced T-B performance of the former as compared to the latter.

The second point concerns the presence of realistic effects in such a terminated-waveguide (focusing) structure. A key such effect is the ubiquitous presence of surface roughness at the media interfaces of the structure. Its detrimental role, particularly with respect to the ability of such a structure to 'trap' / localize a pulse at its end (that is, in creating an open 'cavity' at its end), is well established [9]. For instance, in Fig. S3(a) below, taken from Ref. [9], it is clearly seen that (realistic levels of) surface roughness may completely prevent such a structure from focusing a light pulse at its end/tip, thereby fundamentally undermining its ability to trap a pulse at its end/tip and potentially overcome the T-B limit (note that even if that feat was accomplished, surface roughness would have further added to the already high radiation / back-scattering losses of such a reciprocal structure, as explained above). In contrast, no such deleterious effect arises in the nonreciprocal / topological structure, as it fundamentally allows for robust, one-way propagation, and pulse trapping at the terminating end, unharmed by surface roughness. We show this clearly in Figs. S3(b), (c) below (based on full-wave FDTD simulations we have performed): From Fig. S3(b) one may see that even a large obstacle (metal rod) placed on the way of the guided surface magnetoplasmon leaves the wave unaffected / unperturbed, as the wave smoothly goes around the obstacle and continues its propagation towards the right end without any reflection(s) from the obstacle. In contrast, from Fig. S3(c) it is seen that the same obstacle placed on the way of the guided wave of the reciprocal structure leads, as expected, to significant back-scattering (reflection), i.e. it drastically affects (deteriorates) the ability of such a reciprocal structure to focus light at its end with high efficiency. Thus, not only is the nonreciprocal / topological structure fundamentally superior to its reciprocal / ordinary counterpart in terms of the several directly-T-B-related aspects elucidated above (cf., e.g., Fig. S2 herein) but also, equally importantly, a realistic reciprocal structure often exhibits very poor focusing efficiency or even complete absence of (even temporarily / for brief times) trapped / localized pulses at its end / tip, thereby, essentially, not even being able to 'compete' (in terms of, e.g., overcoming the T-B limit) with its reciprocal / topological counterpart in the first place.

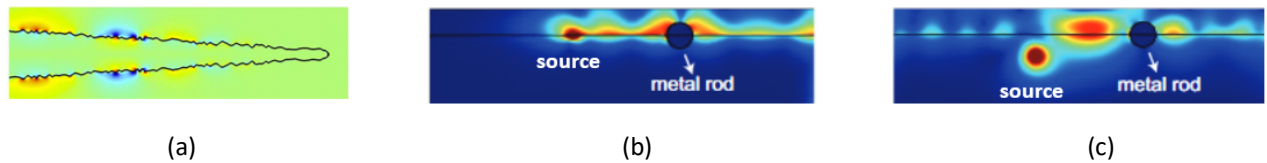

(a)  (b)  (c)

**Fig. S3. Role of realistic material effects on the focusing properties of terminated guides.** (a) Focusing ability of a tapered nanoguide with realistic levels of roughness on its surfaces, where shown is the electric field of the guided surface plasmon supported by the structure. Picture taken from Ref. [9]. (b) A nonreciprocal / topological guide, similar to the one of Ref. [2], where a metallic rod (obstacle) is placed on the way of the excited surface magnetoplasmon. (c) Same as in (b) but in its reciprocal regime of operation.

This focusing inefficiency of (e.g., adiabatically tapered) reciprocal / ordinary guides is, in fact, one of their well-known characteristics / limitations: For instance, typical throughput efficiencies of near-field scanning optical microscopes (NSOMs) – the incumbent technology for breaking the light diffraction limit, and a structure very similar to what is herein discussed – are of the order of, only, $10^{-4}$ to $10^{-5}$ [please see, e.g.: *Nature Photonics* **3**, 220–224 (2009)], and constitute, arguably, the most well-known limitation of this technology. In contrast, as was shown in Fig. S2(c) herein and as was concisely explained above, nonreciprocal / topological guides can exhibit orders-of-magnitude enhanced focusing compared with their reciprocal / ordinary counterparts, unharmed by realistic material effects such as surface roughness, disorders, inhomogeneities, obstacles, material imperfections, etc, owing to their topological nature. Thus, the statement in Ref. [1] that "… *We stress, however, that this broadband focusing is not directly a consequence of nonreciprocity: adiabatically tapered terminated plasmonic waveguides [53–55], which slowly focus the incoming fields toward an apex,* **perform the same function**. …" is indeed not correct as it misrepresents the *fundamental* differences between reciprocal and nonreciprocal cavities.

6) **Role of nonlocality** - the following subtle but important points are to be clarified on this issue:

**First, we note that the objective of Ref. [2], as well as of the present Comment, was to show that the T-B limit can be exceeded, by essentially an arbitrarily high degree, in local (non-spatially-dispersive), linear, time-invariant structures – that is, the same type of structures considered in Ref. [1], as well as in similar previous works [12-14], which reasoned that no such violations may exist in such structures for fundamental reasons [14]. Thus, the new results and physical justifications presented in the main Comment, as well as in Ref. [2], rigorously show that the T-B limit characterizing local, linear, time-invariant structures can be overcome so long as such a violation is *topologically enforced and protected* (see discussion in point-5 above).**

**Second, even when nonlocal effects are considered, one may always redesign the terminated structure considered here and in Ref. [2], e.g. simply by removing the dielectric (Si) layer, such that it can** robustly preserve its unidirectional and topological character **even in the presence of nonlocality**, and for arbitrarily small levels of dissipation, **as has recently been shown in Ref. [11], which also studied terminated one-way waveguides (analogous and very similar to our herein studied one) completely robust to nonlocal effects** – thus, nonlocality cannot for fundamental reasons, i.e. for all possible structures, destroy topological protection (topology), since the latter is a deeper and more fundamental property. In other words, our criticisms of the assertions made in Ref. [1] are, fundamentally, unaffected by the possible presence of nonlocality.

Third, we note that there is no need for termination and its associated large field-enhancement in a tight region (see Fig. 3(c)), which might give rise to nonlocal effects, as ultrabroadband light trapping [15, 16] and releasing [17] can also exist in **topological** (unidirectional) 'trapped rainbow' structures [18, 19], which can stretch out and localize (trap) a lightfield in tapered guides in a manner stable even under fabrication disorders [15].

Finally, forth, for device applications of such T-B violations, other important phenomena, such as nonlinear and thermal effects [19], will instead need to be considered, both of which can however be addressed by e.g. lowering the injected light power or by resorting to cryogenic conditions – all, interesting future objectives but, clearly, outside the scope of our present work.